\documentclass[reprint, superscriptaddress,
 amsmath,amssymb,
 aps,
 prb,
]{revtex4-2}

\usepackage{graphicx}
\usepackage{dcolumn}
\usepackage{bm}
\usepackage{graphicx}
\usepackage{color}
\usepackage[colorlinks, linkcolor=blue, anchorcolor=blue, citecolor=blue, urlcolor=blue]{hyperref}
\usepackage{amsmath}
\usepackage{multirow}
\usepackage{tabularx}
\usepackage{rotating}
\usepackage{upgreek}
\UseRawInputEncoding

\emergencystretch=\maxdimen
\newcommand{\NBGO}{Nd$_2$Be$_2$GeO$_7$}
\newcommand{\PBGO}{Pr$_2$Be$_2$GeO$_7$}
\begin{document}


\title{Discovery of hidden order in the Shastry-Sutherland magnet \NBGO}



\author{Andi Liu}
\affiliation{Songshan Lake Materials Laboratory, Dongguan 523808, China}
\affiliation{School of Physics and Wuhan National High Magnetic Field Center,
Huazhong University of Science and Technology, Wuhan 430074, China}

\author{Samuel H.Moody}
\affiliation{Laboratory for Neutron Scattering and Imaging DMC, PSI, CH-5232 Villigen PSI, Switzerland}

\author{Thomas J. Hicken}
\affiliation{PSI Center for Neutron and Muon Sciences, CH-5232 Villigen PSI, Switzerland}

\author{Jonas A. Krieger}
\affiliation{PSI Center for Neutron and Muon Sciences, CH-5232 Villigen PSI, Switzerland}

\author{Hubertus Luetkens}
\affiliation{PSI Center for Neutron and Muon Sciences, CH-5232 Villigen PSI, Switzerland}

\author{George D. A. Wood}
\affiliation{ISIS Neutron and Muon Source, Rutherford Appleton Laboratory, Chilton, Didcot OX11 0QX, United Kingdom}

\author{Helen C. Walker}
\affiliation{ISIS Neutron and Muon Source, Rutherford Appleton Laboratory, Chilton, Didcot OX11 0QX, United Kingdom}

\author{Zhendong Fu}
\affiliation{Songshan Lake Materials Laboratory, Dongguan 523808, China}

\author{Jason S. Gardner}
\affiliation{Material Science and Technology Division, Oak Ridge National Laboratory, Oak Ridge, Tennessee 37831, USA}

\author{Jinkui Zhao}
\email{jkzhao@gbu.edu.cn}
\affiliation{School of Physical Sciences, Great Bay University, Dongguan 523808, China}
\affiliation{Songshan Lake Materials Laboratory, Dongguan 523808, China}
\affiliation{Institute of Physics, Chinese Academy of Sciences, Beijing 100190, China}

\author{Zhaoming Tian}
\email{tianzhaoming@hust.edu.cn}
\affiliation{School of Physics and Wuhan National High Magnetic Field Center,
Huazhong University of Science and Technology, Wuhan 430074, China}

\author{Hanjie Guo}
\email{hjguo@sslab.org.cn}
\affiliation{Songshan Lake Materials Laboratory, Dongguan 523808, China}


\date{\today}

\begin{abstract}
  Hidden order typically manifests as a thermodynamic phase transition without a conventional order parameter, leaving its true nature concealed even at the lowest temperatures. In the frustrated Shastry-Sutherland magnet \NBGO, we observe a related yet fundamentally distinct phenomenon. A sharp specific-heat anomaly appears at 250~mK, but zero-field neutron diffraction and muon spin relaxation detect no static magnetism down to 100 and 30 mK, respectively, pointing to a hidden-order state. Remarkably, this hidden order does not emerge under an applied magnetic field, but instead reveals itself only after the field is applied and subsequently removed where magnetic Bragg peaks appear, albeit with strongly suppressed moments. A glassy state is ruled out by ac susceptibility and specific heat measurements. Complementary $\mu$SR measurements reveal coherent spin fluctuations at a rate on the order of gigahertz. Taken together, these results suggest that the system lies in close proximity to the quantum spin liquid and long-range magnetic order state such that a small perturbation can effectively drive the system towards distinct ground states. These findings also distinguish \NBGO\ from known frustrated systems, establishing it as a unique platform where the synergistic interplay among the spin-orbit coupling, crystal field, and magnetic frustration leads to unexpected quantum states.
\end{abstract}


\maketitle


In frustrated magnets, competing interactions can suppress conventional magnetic order down to the lowest temperatures, thereby stabilizing a variety of exotic phases \cite{Gardner2010}. Among them, hidden-order phases, characterized by a clear thermodynamic anomaly yet without any detectable dipolar order parameter under conventional probes such as neutron scattering and muon spin relaxation, remain particularly enigmatic \cite{Mydosh2011,Pourovskii2025}. Most hidden-order systems stabilize a ground state that is largely insensitive to perturbations such as magnetic field due to the underlying high-rank multipolar nature, which does not couple, or otherwise only weakly couple to the magnetic field \cite{Pourovskii2025}. Here, we demonstrate that the Shastry-Sutherland magnet \NBGO\ hosts a hidden-order state but exhibits fundamentally different behavior: although it shows a sharp peak in specific heat, the order parameter is revealed only after applying and subsequently removing the field in neutron diffraction measurements. This history-dependent emergence distinguishes \NBGO\ from all previously studied frustrated magnets and offers a unique platform to explore hidden order in a tunable manner.

The Shastry-Sutherland lattice (SSL), consisting of orthogonal dimers in a 2-dimensional network, is one of the paradigms for studying exchange frustration \cite{shastry1981exact,Miyahara1999,Shin2003,Dorier2008,wu2016,Zayed2017,McClarty2017,Marshall2023,Ferrari2024,Liu2024quantum}. Depending on the ratio of the antiferromagnetic intradimer ($J'$) and interdimer ($J$) interactions, $\delta = J/J'$, the SSL system may form diverse states ranging from the spin-dimer singlet state below the critical point of $\delta$ = 0.675, to the plaquette state (PS) between $\delta$ = 0.675 and 0.765, and the long-range antiferromagnetic (AF) order state when $\delta$ is above 0.765 \cite{Corboz2013}. This is well demonstrated by the pressure-tuned phase transitions in SrCu$_2$(BO$_3$)$_2$ \cite{Zayed2017,Shi2022,Cui2023}. Furthermore, recent density-matrix renormalization group calculations show an additional spin liquid (SL) phase between the PS and AF phases \cite{Yang2022}. However, experimental observation of this SL phase remains scarce.

In recent years, rare-earth-based Shastry-Sutherland magnets have been heavily investigated \cite{Bill2022,Pasco2023,Duan2024,Liu2024Theory,Pula2024,Liu2024,Liu20242} since the interplay between spin-orbit coupling (SOC) and crystal-electric-field (CEF) effect may lead to significant anisotropy and intriguing states such as a quantum entangled triplet ground state. The RE$_2$Be$_2$GeO$_7$ (RE = rare-earth elements) series of compounds is of particular interest since nearly all the rare-earth elements can be incorporated into the lattice, and the structure is free from antisite disorder \cite{Liu2024,Liu20242}, making it an ideal platform for investigating emergent phases as a function of anisotropy, pressure and magnetic field.

\begin{figure*}
\centering
\includegraphics[width=2\columnwidth]{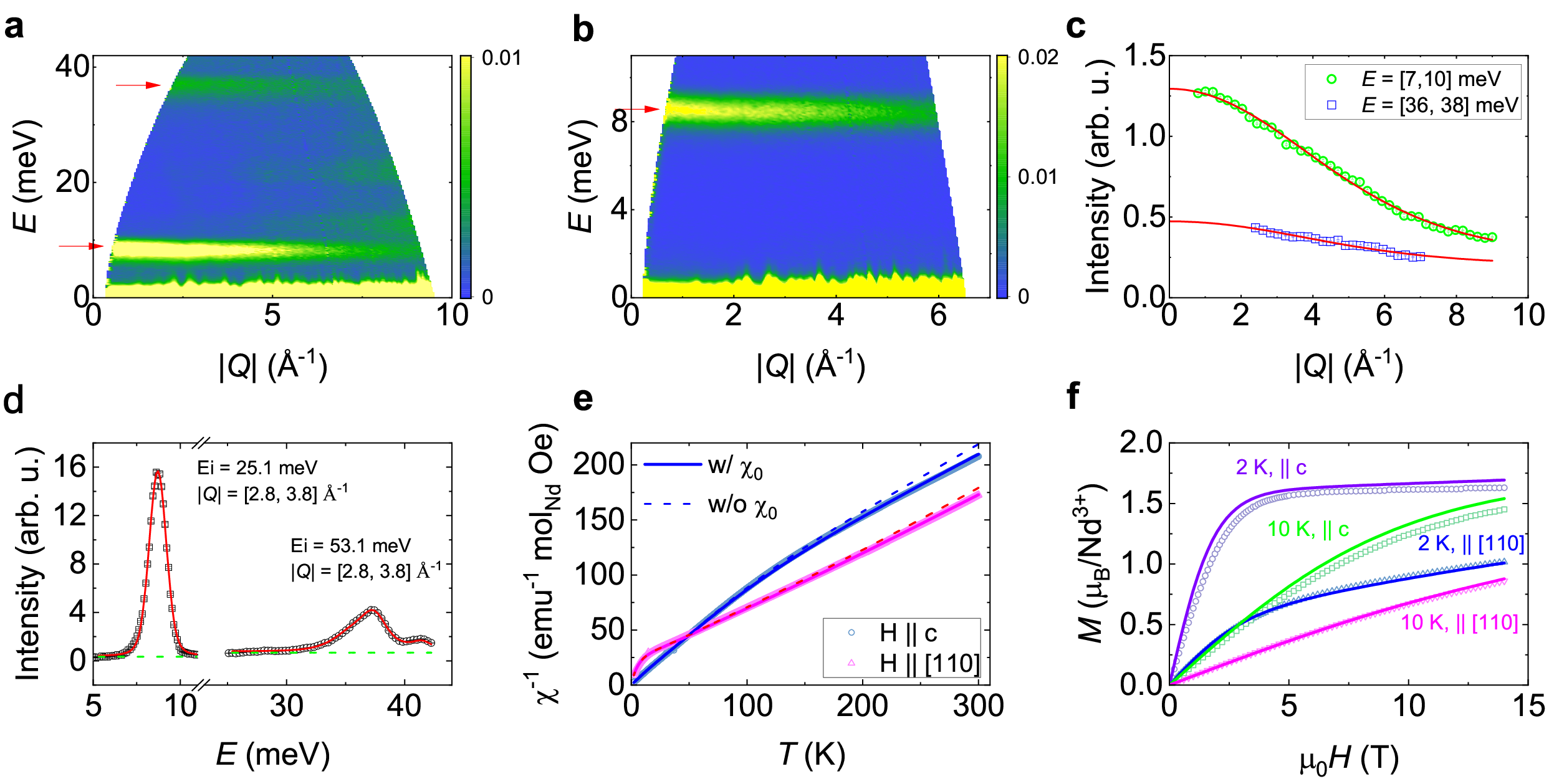}
\caption{\textbf{Crystal electric field excitations.} \textbf{a-b}, Inelastic neutron scattering intensity maps collected with incident neutron energies $E_i$ of 53.1 and 25.1~meV, respectively, at 5~K. The red arrows indicate the CEF excitations. \textbf{c,} Momentum transfer ($|Q|$) dependence of the integrated intensities (\textit{I}) for the excitations at around 9 and 37~meV. The red curves are fits according to the form factor of Nd$^{3+}$, $I = I_0f^2(Q) + I_{bg}$, where $I_0$ and $I_{bg}$ are fitting parameters. \textbf{d,} Energy ($E$) dependence of the intensities integrated over $|Q|$ along with a CEF fit. The green dashed lines are backgrounds. The error bars in \textbf{c-d} correspond to statistical errors equal to the square root of the intensities. \textbf{e,} Temperature dependence of the inverse magnetic susceptibility with magnetic field applied along the \textit{c}- and [110] directions. The solid and dashed lines are fits according to the CEF model with and without a small constant term $\chi_0$, respectively. \textbf{f,} Isothermal magnetizations measured along the \textit{c}- and [110] directions at various temperatures. The solid curves are calculated according to the best-fit CEF model.}
\label{INS}
\end{figure*}

\noindent
\textbf{Results}

\noindent
Earlier studies on Nd$_2$Be$_2$GeO$_7$ \cite{Liu20242} unveiled a steep decrease in the magnetic susceptibility below $\sim$0.4~K. The zero-field specific-heat data further exhibit a broad peak at about 0.5~K, followed by a sharp, $\lambda$-shaped peak at about 0.25~K, indicating a long-range magnetically ordered state.
The rare-earth magnetic properties are heavily influenced by the CEF effect. To better understand the CEF scheme of Nd$^{3+}$ ions in \NBGO, we performed inelastic neutron scattering (INS) measurements at 5 K (paramagnetic state). The INS spectrum in Fig.~\ref{INS}a shows two dispersionless excitations at $\sim$8.7 and 37.5~meV. The momentum transfer ($|Q|$) dependence of the integrated intensities for these two excitations follows the form factor of Nd$^{3+}$ (Fig.~\ref{INS}c), suggesting their magnetic origin.
A closer inspection of the data reveals two additional peaks at $\sim$36 and 41.5~meV; see the Supplemental Materials (SM) for more details \cite{SM}. These four excitations are consistent with CEF splittings expected for Nd$^{3+}$ ions with \textit{J} = 9/2.

\begin{figure*}
\centering
\includegraphics[width=2\columnwidth]{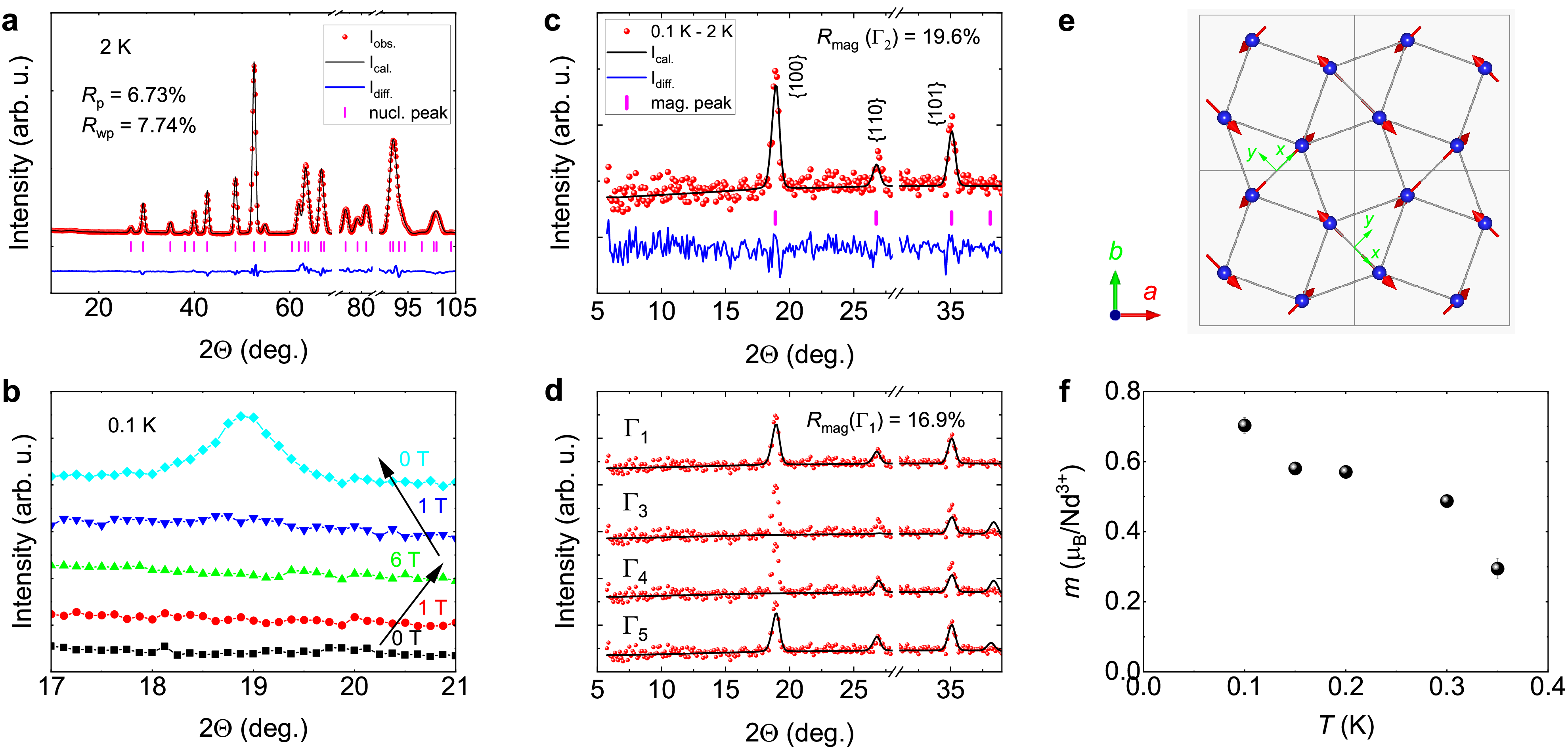}
\caption{\textbf{Neutron diffraction measurements.} \textbf{a,} Neutron powder diffraction pattern for \NBGO\ collected at 2~K, along with a Rietveld refinement based on the crystal structure with the space group $P\bar{4}2_1m$. The excluded regions correspond to the peaks from the copper holder and the aluminum from the cryostat. \textbf{b,} Evolution of the low-angle intensities under a magnetic field cycled from 0~T to 6~T and back to 0~T at 0.1 K. \textbf{c,} Magnetic structure refinement based on the irreducible representation (IR) $\Gamma_2$ against the difference pattern ($I_\mathrm{0.1\,K} - I_\mathrm{2\,K})$. \textbf{d,} Representative refinements based on other IRs. \textbf{e,} Magnetic structure for \NBGO\ based on the best refinement according to $\Gamma_2$. The red and green arrows within the lattice represent the moments and local coordinates, respectively. \textbf{f,} Temperature dependence of the moment size for \NBGO. The error bars are standard errors from the fits.}
\label{NPD}
\end{figure*}

To determine the CEF scheme, we analyzed these spectra based on a weak coupling scheme acting on the $|J,m_J\rangle$ basis using the PyCrystalField package \cite{Scheie2021}.
The CEF Hamiltonian can be written as $\mathcal{H}_\mathrm{CEF} = \sum B_l^mO_l^m$,
where $O_l^m$ are the Stevens operators and $B_l^m$ are the CEF parameters \cite{Stevens1952,Hutchings1964}. With the quantization axes along the \textit{c} axis, the $C_s$ point symmetry for the Nd$^{3+}$ ions at the $4e$ site allows up to 15 nonzero $B_l^m$.
The limited number of observables in neutron scattering may result in a certain degree of ambiguity in determining the CEF parameters. To overcome this obstacle, the CEF model is fitted simultaneously to the temperature-dependent susceptibility along different crystallographic directions. The best fits are shown in Fig.~\ref{INS}d-e. Based on the best-fit CEF model, the calculated isothermal magnetizations agree reasonably well with the experimental data, as shown in Fig.~\ref{INS}f. The obtained CEF eigenvalues and eigenvectors can be found in the SM Tab. S1 \cite{SM}.
From the obtained \textit{g} tensor for the CEF ground state, the corresponding \textit{g} factors are $g_x$~=~2.02, $g_y$~=~0.14 and $g_z$~=~3.18, where the $y$ direction is perpendicular to the local mirror plane, the $z$ direction is along the crystallographic \textit{c} axis, and the $x$ direction is perpendicular to the \textit{yz} plane; see also Fig.~\ref{NPD}e. The $g_z$ value is in good agreement with that determined by electron spin resonance (ESR) measurements ($g_z^{\mathrm{ESR}}$ = 3.1) \cite{SM}. The ground state ordered moments, computed from $g_J\mu_\mathrm{B}$$\langle J_\alpha \rangle$, where $g_J$ is the Land\'{e} \textit{g} factor, are $m_x$ = 0.82 $\mu_\mathrm{B}$, $m_y$ = 0, and $m_z$ = 1.55 $\mu_\mathrm{B}$. These results demonstrate that the moments are constrained within the mirror plane in the paramagnetic state.

Having established the CEF scheme, we move to the neutron diffraction measurements. Figure \ref{NPD}a shows the neutron powder diffraction pattern measured in the paramagnetic state (2~K), which can be well described by the tetragonal crystal structure as reported before \cite{Liu20242}.
Intriguingly, when the temperature is lowered to 0.1~K, no additional reflections, nor any enhancement of the nuclear peaks could be observed (see Fig.~\ref{NPD}b), suggesting the absence of magnetic ordering. This is corroborated by zero-field (ZF) muon spin relaxation ($\mu$SR) measurements, as will be shown later. The absence of a long-range magnetic order in neutron diffraction and $\mu$SR, concomitant with a sharp specific-heat anomaly is reminiscent of a hidden-order phase observed in many other systems \cite{Cox1967,Palstra1985}. When a magnetic field was applied, no discernible change in neutron diffraction could be observed up to 6~T. Strikingly, when the field was ramped down to 0~T, additional reflections appear, as shown in Fig.~\ref{NPD}b.
After subtracting the 2-K data from the low-temperature one, three magnetic reflections can be identified and indexed by a propagation vector \textbf{k}~=~(0,~0,~0); see Fig.~\ref{NPD}c. To determine the magnetic structure, we have performed irreducible representation (IR) analysis showing that the reducible representation for the Nd$^{3+}$ ions at the $4e$ site is decomposed as $\Gamma_\mathrm{mag} = \Gamma_1 + 2\Gamma_2 + 2\Gamma_3 + \Gamma_4 + 3\Gamma_5$.
The basis vectors for each IR can be found in the SM Tab. S2 \cite{SM}. As shown in Fig.~\ref{NPD}c-d, only the models based on $\Gamma_1$ and $\Gamma_2$ can describe the experimental data satisfactorily. The magnetic moments are pointing out of the mirror plane for $\Gamma_1$, whereas they are constrained within the mirror plane for $\Gamma_2$. For these two models, only the latter is consistent with the bulk magnetization measurements \cite{Liu20242} and CEF analysis. The corresponding magnetic structure is shown in Fig.~\ref{NPD}e. Note that the spins are non-collinear, in contrast to that observed in SrCu$_2$(BO$_3$)$_2$ \cite{Cui2023}. This structure is also different from that in BaNd$_2$ZnO$_5$ and BaNd$_2$ZnS$_5$ where ferromagnetic dimers are formed \cite{Ishii2021,Marshall2023}. Our ESR measurements indicate the presence of Dzyaloshinskii-Moriya interactions between the dimer spins, which may account for the non-collinear structure \cite{SM}.
From the peak widths (half width at half maximum) of the \{1~0~0\} and \{1~0~1\} peaks, we estimate the correlation length $\xi$ of 89(6) and 94(7) \AA, respectively.
The temperature dependence of the moment size ($m$) is depicted in Fig.~\ref{NPD}f. The ordered moment vanishes above $\sim$0.35~K. This temperature scale matches that observed in the magnetic susceptibility and the broad hump rather than the sharp peak seen in the specific heat \cite{Liu2024}, possibly due to residual inelastic signal in the diffraction channel. The $m$ value at 0.1~K amounts to 0.60(2)~$\mu_\mathrm{B}$ and 0.36(3)~$\mu_\mathrm{B}$ along the local $x$ and $z$ directions, respectively. The $m_x$ component agrees reasonably well with that expected for the CEF ground state (0.85~$\mu_\mathrm{B}$). On the contrary, the $m_z$ component is much reduced compared to 1.55~$\mu_\mathrm{B}$, indicating much stronger quantum fluctuations along the \textit{c} axis.

\begin{figure*}
\centering
\includegraphics[width=2\columnwidth]{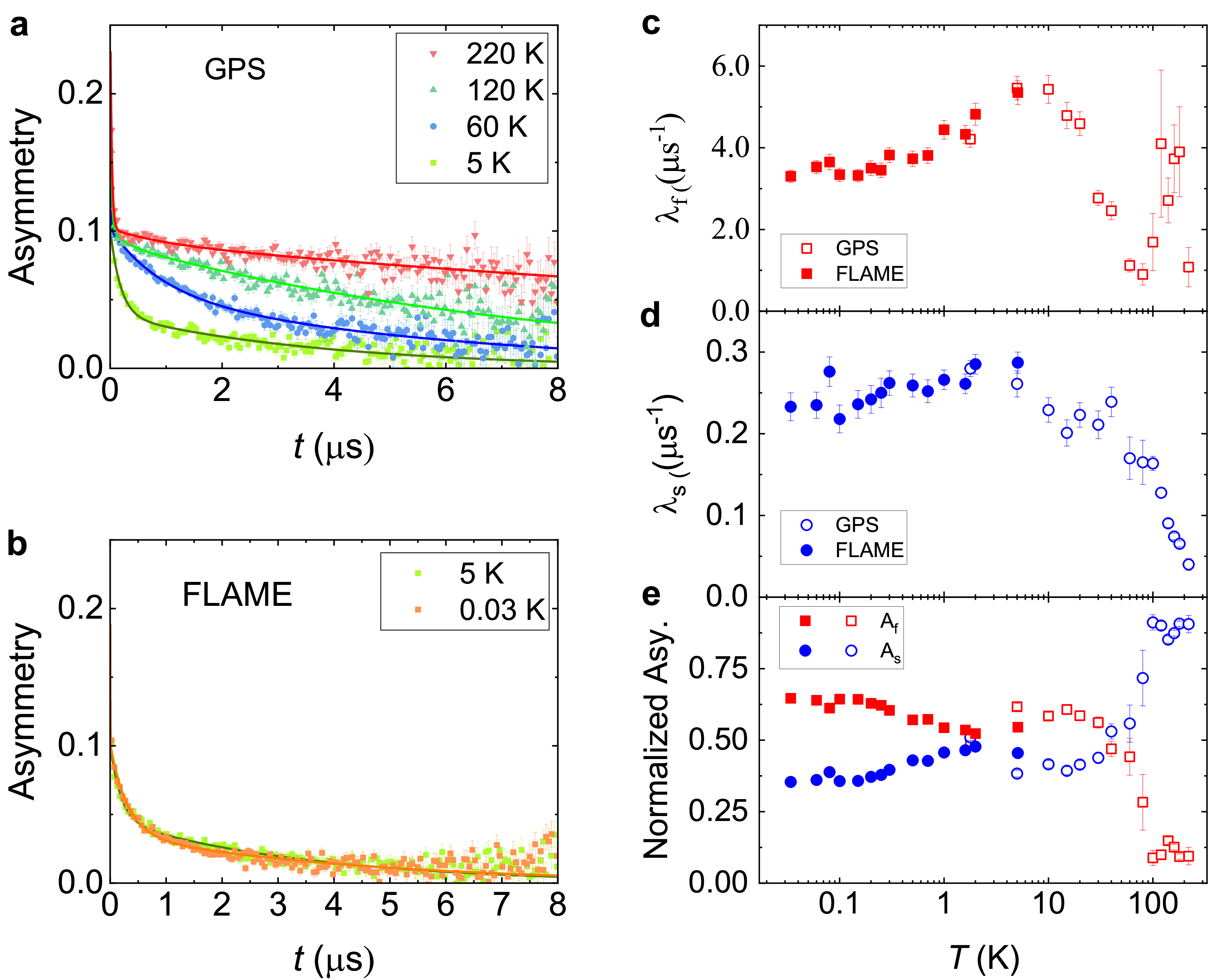}
\caption{\textbf{Zero-field $\mu$SR measurements.} \textbf{a-b,} Representative zero-field $\mu$SR spectra measured at various temperatures on the GPS and FLAME spectrometers, respectively. The solid curves are fits according to Eq. \ref{eq2}. \textbf{c-d,} Temperature dependence of the fast and slow muon spin relaxation rate $\lambda_f$ and $\lambda_s$. \textbf{e,} Temperature dependence of the amplitudes $A_f$ and $A_s$. The error bars in \textbf{a-b} and \textbf{c-e} correspond to statistical errors and standard errors from the
fits, respectively.}
\label{muSR}
\end{figure*}

\begin{figure}
\centering
\includegraphics[width=1\columnwidth]{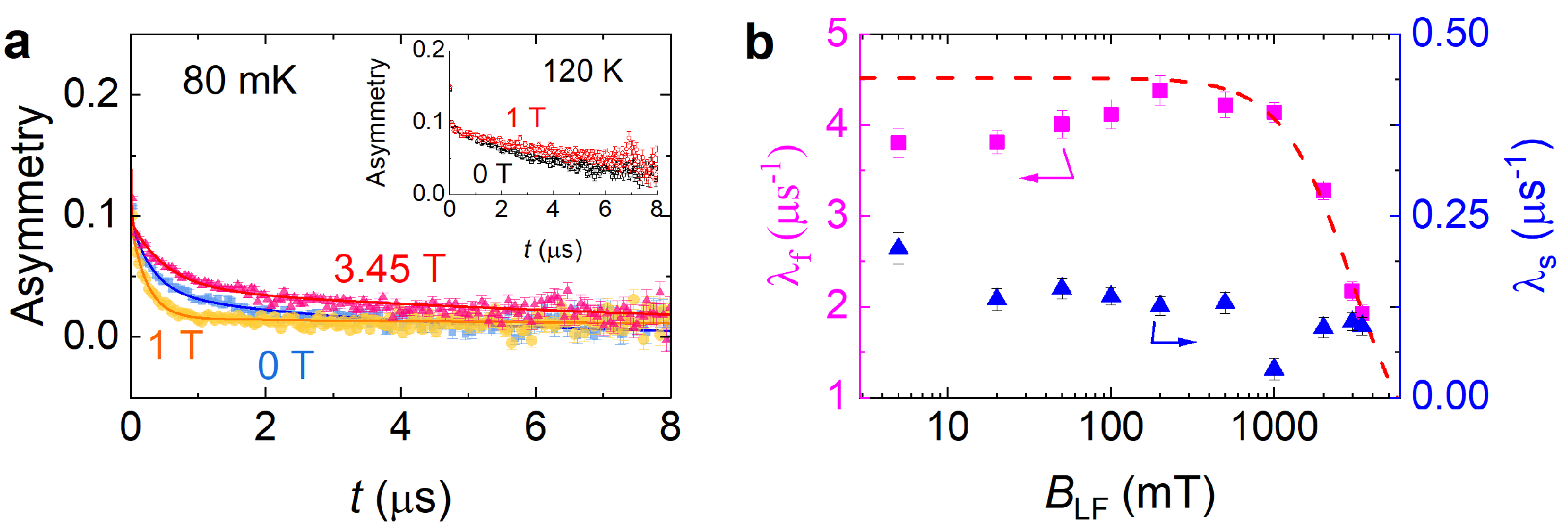}
\caption{\textbf{Longitudinal-field $\mu$SR measurements.} \textbf{a,} Representative $\mu$SR spectra measured at various external magnetic fields at 80~mK. The solid curves are fits according to Eq.~\ref{eq2}. The inset shows the LF measurements at 120~K. The error bars are statistical errors. \textbf{b,} Temperature dependence of the relaxation rates $\lambda_f$ and $\lambda_s$. The error bars are standard errors of the fits. The red curve is a fit to $\lambda_f$ as described in the text.}
\label{LF}
\end{figure}

To further probe the magnetic ground state, we employed the $\mu$SR technique, which is sensitive to magnetic fields down to $\sim$10$^{-2}$ mT \cite{Amato2024}. Some typical ZF $\mu$SR spectra measured on the GPS and FLAME spectrometers at the Paul Scherrer Institut are shown in Fig.~\ref{muSR}a and \ref{muSR}b, respectively. For both measurements, the asymmetry decreases sharply within 0.05 $\mu$s, and decays slowly afterwards. The early-time fast relaxation persists up to 220~K, as well as in a longitudinal field (LF) of 3.45~T; see Fig. \ref{LF}a and SM Fig. S1. The origin of this component remains unclear; however, it cannot arise from magnetic ordering, as no long-range ordered state is expected at 220~K. Next, we focus on the relatively slowly relaxing component. This component decays faster with decreasing temperatures, and tends to remain unchanged below about 5~K. No spontaneous muon spin precession is observed down to 30~mK, indicating the absence of long-range magnetic ordering, consistent with the NPD measurements. Even after a field-training process, i.e., increasing the field up to 3.45 T and ramping back down to zero field at 80 mK, no oscillation is present. This is in contrast to the field-training results observed in the NPD experiment. The maximum field in the $\mu$SR experiment is smaller than that of the neutron experiment. However, we note that the field strength is already much larger than the magnetization saturation field, which is about 2 T at 0.5 K when the field is applied along the \textit{c} axis; see the SM \cite{SM}. Therefore, the seemly contradictory results from these two techniques could originate from the different time scales they have probed, as will be verified by our LF $\mu$SR measurements, reflecting the dynamic nature of the ground state.

For a quantitative description of the spectra, the asymmetry is fitted by a three-component function:
\begin{equation}\label{eq2}
  A(t) = A_1\mathrm{exp}(-\lambda_1 t) + A_f\mathrm{exp}(-\lambda_f t) + A_s\mathrm{exp}(-\lambda_s t),
\end{equation}
where the first term takes into account the early-time fast relaxation. The amplitude of $A_1$ is fixed to the value determined at the highest measured temperatures, with $A_1/A(0)$ being 37\% and 42\% for the GPS and FLAME datasets, respectively. The small discrepancy could arise from a slight uncertainty in determining the initial asymmetry for the two spectrometers. More analyses of this term can be found in \cite{SM}. The remaining two terms reflect a spatially distinguishable regions where the relaxation rates are fast (\textit{f}) and slow (\textit{s}), respectively.
The temperature dependence of the extracted parameters are shown in Fig.~\ref{muSR}c-e. As can be seen, the results obtained from the two spectrometers are in good agreement. The fast relaxation rate, $\lambda_f$ shows an anomalous temperature dependence. It exhibits a peak at about 180 K, and starts to increase below $\sim$80 K. At $\sim$10 K, it shows another broad peak, and then becomes almost temperature independent below 1 K.
We note that the amplitude of $A_f$ is very small above 100~K, resulting in a large uncertainty in the fast relaxation rate. The peak at 180~K is therefore more like an artefact.

The increase below 80 K reflects a slowing down of the internal fields. The temperature independent behavior at low temperatures has been observed in many frustrated systems \cite{Arh2022,Yadav2025,Cao2025}, indicating the persistence of spin fluctuations. The slow relaxation rate, $\lambda_s$ is about one order of magnitude smaller than $\lambda_f$. It also exhibits persistent fluctuations below $\sim$1 K, while the high temperature anomalies are less pronounced than those observed in $\lambda_f$. The temperature dependence of the amplitudes, $A_f$ and $A_s$, suggests that the regions of fast relaxation rate become the dominance at low temperatures.

The unusual response to magnetic fields for \NBGO\ as observed in NPD implies that it may host a peculiar spin dynamics. This is indeed verified by our LF $\mu$SR measurements performed at 80~mK. In general, the muon spins will be gradually decoupled from the local fields with increasing LF, resulting in a decrease in the muon spin relaxation rate. However, the 80-mK spectrum measured at 1~T exhibits a faster relaxation compared to that at 0~T; see Fig.~\ref{LF}a. Only at higher fields does the spectra behave as expected. Such a field-enhanced relaxation is reminiscent of the avoided level crossing (ALC) arising from the quadrupolar splitting of the nuclear moments with spin  $I >$ 1/2 \cite{Kreitzman1986,Cox1992,Jiang2025}. However, one would expect similar resonance behavior to persist at high temperatures, as observed in Cu \cite{Cox1992} and Ba$_6$Nd$_2$Ti$_4$O$_{17}$ \cite{Jiang2025}. Such behavior is absent in \NBGO\ (see the inset of Fig.~\ref{LF}a), suggesting instead that magnetic fluctuations related to the electronic spins play a vital role.
The LF spectra can also be described by Eq.~\ref{eq2}. The extracted LF dependence of the relaxation rates is shown in Fig.~\ref{LF}b. While $\lambda_s$ is gradually suppressed by LF, $\lambda_f$ increases firstly, and shows a broad maximum at  $\sim$0.5~T.
The LF $\mu$SR measurements probe the spectral density $J(\omega)$ at the Larmor frequency $\omega_\mu = \gamma_\mu B_\mathrm{LF}$ \cite{Amato2024}, where $\gamma_\mu/2\pi = 135.5$~MHz/T is the gyromagnetic ratio of the muon. $J(\omega)$ is the Fourier transform of the field autocorrelation function $\langle\delta B(0)\delta B(t)\rangle$ \cite{Amato2024}. In the fast-fluctuation limit, the relaxation rate is expressed by the Redfield formula:
\begin{equation}\label{eq5}
  \lambda(\omega_\mu) = \gamma_\mu^2 J(\omega_\mu) = \frac{2(\gamma_\mu \Delta)^2\nu}{\nu^2+\omega_\mu^2},
\end{equation}
where $\nu$ is the fluctuation rate of the internal fields, and $\Delta$ is the field distribution width. A fit of Eq.~\ref{eq5} to the high-field data above 0.5~T yields $\nu$ of 2.6(1) GHz and $\Delta$ of 894(16) G, so that $\nu/\gamma_\mu\Delta$ = 34, fulfilling the fast-fluctuation limit. The discrepancy between the $\mu$SR and NPD measurements can be reconciled by considering that the latter has a time scale of $\sim$10$^{-10}$ s \cite{Dalmas2006}, which is smaller or comparable with the spin-spin correlation time 1/$\nu$ as inferred from the LF $\mu$SR measurements. Therefore, although the ground state is highly dynamic, the NPD results imply that the spins are fluctuating coherently in the 10$^{-10}$ s time scale, which is in the fast fluctuation regime for the slow $\mu$SR technique.

\noindent
\textbf{Discussion}

\noindent
The observed field response for \NBGO\ appears inconsistent with a high-rank multipolar order parameter as proposed for URu$_2$Si$_2$ \cite{Mydosh2011}. Our detailed ac susceptibility and specific heat measurements also rule out a spin glass ground state \cite{SM}. Alternatively, the observed behavior may instead originate from frustration, which could lead to a highly degenerate ground state, potentially hosting a hidden-order phase. Hidden-order-like phase transitions have been observed in magnetically frustrated systems such as pyrochlore oxides Yb$_2$Ti$_2$O$_7$ \cite{Hodges2002} and Tb$_2$Sn$_2$O$_7$ \cite{Dalmas2006}. In Yb$_2$Ti$_2$O$_7$, $\mu$SR measurements revealed a first-order transition in spin dynamics with the fluctuating rate dropping from GHz to MHz below $\sim$0.24 K. Meanwhile, no ordered signal was picked up in NPD \cite{Hodges2002}. In Tb$_2$Sn$_2$O$_7$, the order was detected both in specific heat and NPD measurements, but not in $\mu$SR measurements, suggesting that the moments fluctuate in the fast limit for $\mu$SR \cite{Dalmas2006}.
The \NBGO\ is, however, also distinct from these known frustrated systems. One possible scenario is that the system explores the landscape of the highly degenerate states and a particular one is stabilized when a magnetic field is applied. Nevertheless, the stabilized phase can only manifest itself in a zero field since the magnetic field would suppress a true antiferromagnetic state.
Once the phase is manifested at zero field, it resembles the ground state of Tb$_2$Sn$_2$O$_7$ \cite{Dalmas2006} in several aspects. For both phases, magnetic reflections are observed, but no long-range magnetic ordering is inferred from $\mu$SR. Moreover, both show an enhanced muon spin relaxation rate at low fields, indicating an enhanced density of excitations. These similarities suggest common physics between these two frustrated systems, although their lattice geometries are rather different. The uniqueness of \NBGO\ lies in the way it reacts to the field which, to the best of our knowledge, has not been reported in other systems. The stabilized AF phase out of a disordered phase in zero field also suggests that the system may reside on the boundary between the SL and AF phases. These distinct characteristics establish \NBGO\ as a benchmark material in the family of Shastry-Sutherland magnets. Future single crystal diffuse scattering, and low-energy inelastic neutron scattering measurements will be highly desirable to extract the inter- and intra-dimer interactions and thereby determine the compound's location within the phase diagram.

In conclusion, we have unveiled a hidden-order phase in the Shastry-Sutherland antiferromagnet \NBGO\ through thermodynamic, neutron-scattering, and muon-spin-relaxation measurements. The ordered phase can be stabilized by magnetic fields, yet the ordered moments along the \textit{c} axis are strongly suppressed relative to those expected from the CEF ground state. A comparison between the neutron and muon results further indicates that the moments fluctuate coherently at gigahertz frequencies. Taken together, the unusual field response of \NBGO\ demonstrates how the synergistic interplay among SOC, CEF effects, and magnetic frustration can stabilize unexpected quantum states.

\bigskip

\noindent
\textbf{Methods}

\noindent
\textbf{Sample preparation} Polycrystalline samples were synthesized by the solid-state reaction method as described in \cite{ashtar2021}. Starting materials of Nd$_{2}$O$_{3}$ (99.99\%), GeO$_{2}$ (99.99\%) and BeO (99.9\%) were mixed in the stoichiometric ratio, ground thoroughly, and heated at 1250$^\circ$C with several intermediate grindings. Single crystals of \NBGO\ were grown by the high-temperature flux method as described in the previous work \cite{Liu2024,Liu20242}. Clean purple single crystals with well-defined facets were successfully grown and mechanically separated from the bulk.

\noindent
\textbf{Magnetization measurements} were performed on a physical property measurement system (PPMS) equipped with the vibrating sample magnetometer (VSM) option in the temperature range of 1.8 $-$ 300 K. The single crystals were oriented with their crystallographic \textit{c} axis and [1~1~0] direction along the magnetic fields.

\noindent
\textbf{$\mu$SR measurements} were performed on the FLAME and GPS spectrometers at the Paul Scherrer Institut (PSI), Villigen, Switzerland. The powder samples were mixed with diluted GE varnish and pressed into pellets of 12~mm in diameter to ensure good thermal contacts at low temperatures. The experimental asymmetry is defined as ${A(t)}=\frac{\alpha F(t)- B(t)}{\alpha F(t)+ B(t)}$, where $F(t)$ and $B(t)$ are the number of positrons arriving at the forward and backward detectors at time $t$, respectively. $\alpha$ reflects the counting efficiency of different detectors and the sample shape. The data were analyzed using the MUSRFIT software package \cite{Suter2012}.

\noindent
\textbf{INS measurements} were carried out on the MERLIN spectrometer at ISIS, UK. The powder samples were loaded into aluminium foil sachets, which were then wrapped around the inner part of a cylindrical aluminium can and cooled down to $\sim$5 K by a closed-cycle refrigerator. MERLIN was operated in multirep mode  with incident neutron energies of 25.1, 53.1 and 179 meV. The data were processed using Mantid \cite{mantid}. The powder averaged neutron cross section with dipole approximation can be expressed as \cite{Boothroyd}
\begin{equation}\label{}
\begin{split}
  \frac{d\sigma^2}{d\Omega d\omega} = & N(\gamma r_0)^2\frac{k_f}{k_i}F^2(Q)\mathrm{exp}[-2W(Q)]\sum_{i,j} p_i\times \\ &\frac{2}{3}\sum_\alpha |\langle \Gamma_j|\hat{J}_\alpha|\Gamma_i\rangle|^2\delta(\hbar\omega + E_i - E_j),
\end{split}
\end{equation}
where $\alpha$ = \textit{x}, \textit{y} and \textit{z} are the local coordinate directions as shown in Fig. \ref{NPD}e, and the other symbols have generic meanings. The Debye-Waller factor, $W(Q)$, was fixed to zero. The magnetization was calculated non-perturbatively by adding a Zeeman term to the CEF Hamiltonian \cite{Scheie2021}. There are two sets of dimers running along and perpendicular to the global [1~1~0] direction, respectively. Thus, the calculated $M_{110}$ = ($M_x + M_y$)/2.

\noindent
\textbf{NPD measurements} were performed using the SINQ cold neutron diffractometer DMC at PSI. The neutron wavelength was 2.454 {\AA}. About 5 g of \NBGO\ pellets were loaded into a cylindrical copper can of 7~mm in diameter and sealed in a helium atmosphere. The data were collected at temperatures ranging from 0.1 to 2 K in different magnetic fields from 0 to 6 T. The (magnetic) structure refinements were carried out using the FullProf software suite \cite{Fullprof}.

\bigskip

\noindent
\textbf{Data availability}

\noindent
The data that support the findings of this study are available from the corresponding authors upon reasonable request.

\bibliography{NBGO}

\bigskip

\noindent
\textbf{Acknowledgements}

\noindent
We thank Wei Li and Changle Liu for valuable discussions. This work was supported by the Guangdong Basic and Applied Basic Research Foundation (Grant No. 2022B1515120020). We gratefully acknowledge the Science and Technology Facilities Council (STFC) for Xpress access to neutron beam time on MERLIN at ISIS (RB2490300, https://doi.org/10.5286/ISIS.E.RB2490300-1). We thank Viviane Pecanha-Antonio for the assistance of INS measurements at MERLIN. Part of this work was based on experiments performed at the Swiss Muon Source S$\mu$S and SINQ, Paul Scherrer Institut, Villigen, Switzerland.

\bigskip

\noindent
\textbf{Author contributions}

\noindent
H.G., Z.T. and J. Z. conceived the project. A. L., H. G. and S. M. performed the neutron diffraction measurements. A. L., H. G., T. H., J. K., H. L. and J. G. performed the $\mu$SR measurements. G. W., V. P. and H. W. performed the inelastic neutron scattering measurements. A. L. and Z. F. conducted the magnetization and specific heat measurements. H. G. analyzed the data. A. L., Z. T., J. Z. and H. G. wrote the manuscript with inputs from all authors.

\bigskip

\noindent
\textbf{Competing interests}

\noindent
The authors declare no competing interests.

\clearpage
\newpage

\renewcommand{\thefigure}{S\arabic{figure}}
\renewcommand{\thetable}{S\arabic{table}}
\setcounter{figure}{0}
\setcounter{table}{0}

\onecolumngrid
\begin{center}
    {\large \bfseries Supplemental Materials for ``Discovery of hidden order in the Shastry-Sutherland magnet \NBGO''}
\end{center}
\vspace{1cm}

\twocolumngrid

\begin{figure*}[t]
\includegraphics[width=0.95\textwidth]{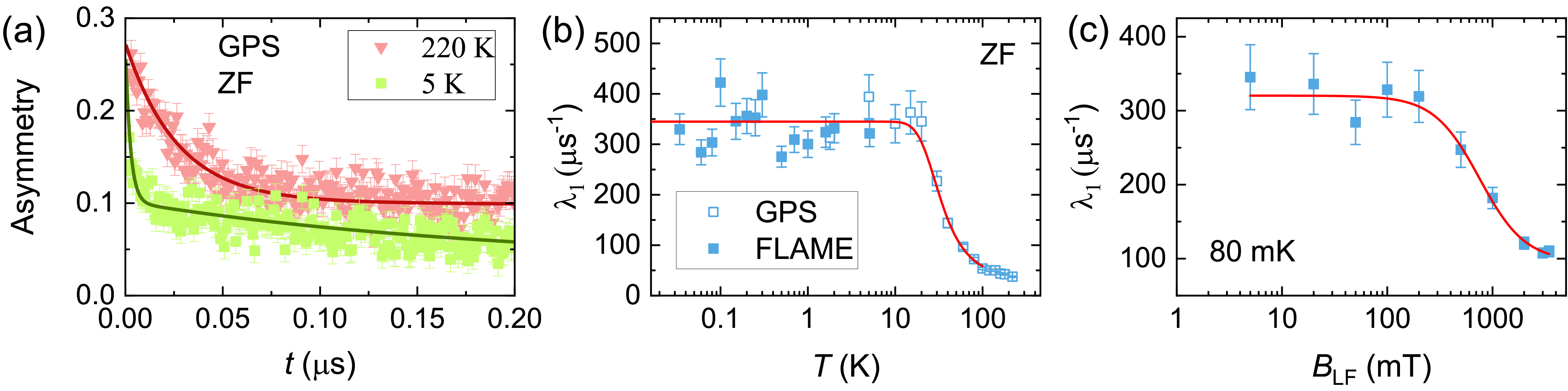}
\caption{(a) Early-time region of the ZF-$\mu$SR spectra collected at GPS. (b) Temperature dependence of the relaxation rate $\lambda_1$; see Eq. (1) in the main text. (c) Longitudinal-field dependence of $\lambda_1$. The red curves are fits; see the text for details.}
\label{gps}
\end{figure*}

\section{$\mu$SR}
Figure \ref{gps}(a) highlights the early-time region of the GPS data. The fast-decay component also exhibits a significant temperature dependence below 220~ K, and the relaxation rate becomes nearly constant below about 20~K; see Fig. \ref{gps}(b). At high temperatures, the muon spin polarization could be relaxed via the Orbach process involving the excited CEF levels \cite{orbach1961}. Such behavior can be phenomenologically described by $\lambda^{-1} = \lambda_0^{-1} + C \mathrm{exp}(-\delta/k_B T)$ \cite{Lago2007,Khasanov2008}, as shown in Fig. \ref{gps}(b). The best fit yields a gap of $\delta$ = 7.8(4) meV, which is in reasonable agreement with the value (8.7~meV) obtained from INS  measurement.

In contrast to the $\lambda_f$ component, $\lambda_1(B_\mathrm{LF})$ follows a modified Redfield theory such that
\begin{equation}
\lambda(B_\mathrm{LF})=\frac{2(\gamma_\mu\Delta)^2\nu}{\nu^2+(\gamma_{\mu} B_\mathrm{LF})^2} + \lambda_0,
\label{Redfield}
\end{equation}
where $\Delta$ is the rms of the internal field distribution width. The deduced parameters are $\nu$ = 0.64(6) GHz, $\Delta$ = 3151(155) G, and $\lambda_0$ = 96(4) $\mu s^{-1}$. The origin of this $\lambda_1$ component is still unclear. It could arise from a muon site close to the magnetic moments such that a large internal field distribution is detected by the muon, or a possible formation of the muonium state. Future muon site calculations, and measurements on a nonmagnetic analogue such as La$_2$Be$_2$GeO$_7$ will be helpful to address this issue.

\section{specific heat}

\begin{figure}
\includegraphics[width=0.45\textwidth]{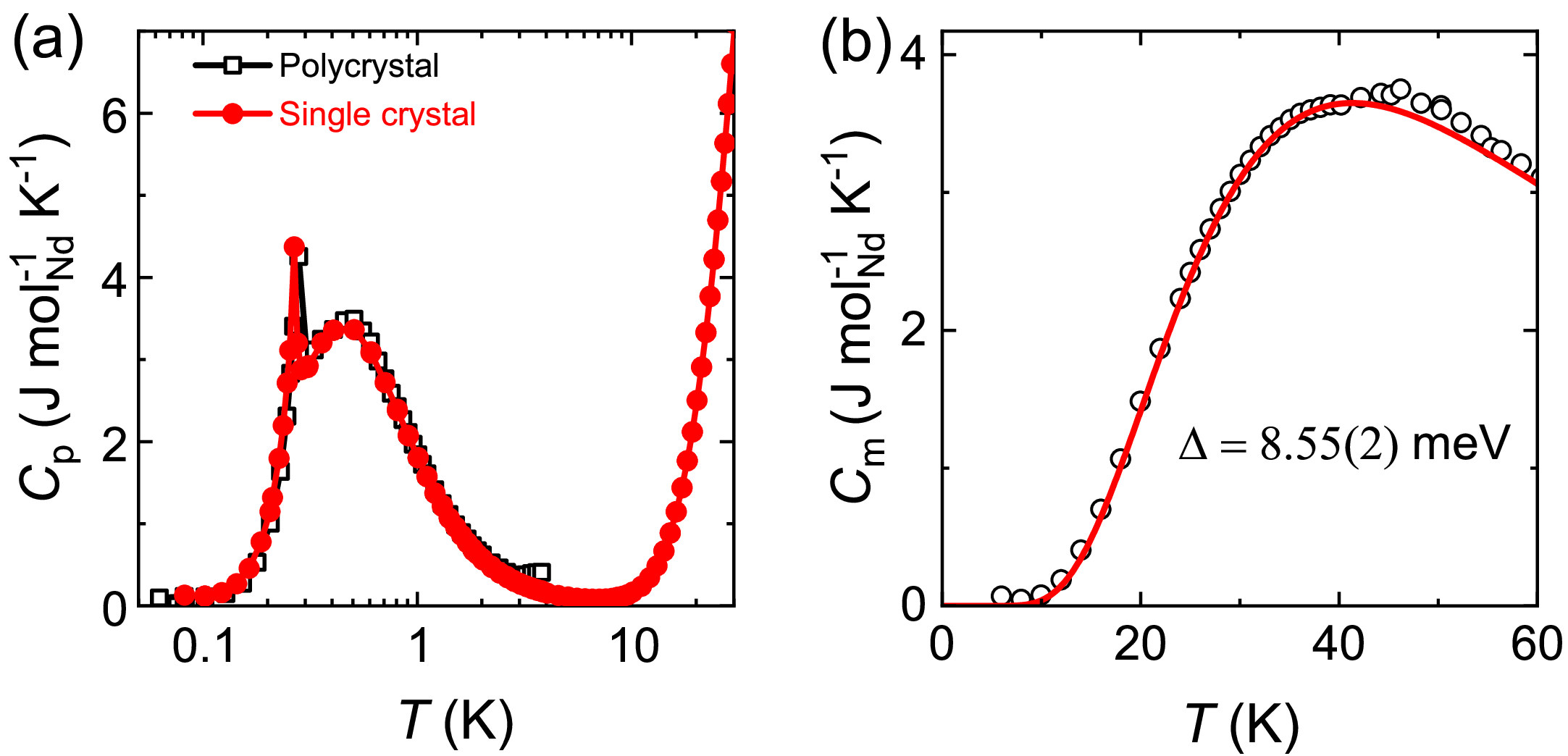}
\caption{Specific heat. (a) Zero-field specific heat of \NBGO\ measured on polycrystalline and single-crystalline samples. (b) Magnetic specific heat for \NBGO\ together with a two-level Schottky fit.}
\label{Cp}
\end{figure}

Polycrystalline and single-crystalline samples may show distinct properties due to subtle variations of the crystal structure, as observed in the pyrochlore oxide Yb$_2$Ti$_2$O$_7$ \cite{Yaouanc2011,Ross2012}. In Fig. \ref{Cp}(a), we present zero-field specific heat measurements on both the polycrystalline and single-crytalline samples of \NBGO. Both measurements show consistent results exhibiting a broad hump at about 0.5~K, followed by a sharp peak at $\sim$0.25 K.

The first CEF excitation at $\sim$8.7~meV should be detectable by specific heat measurements. Indeed, after subtracting the phonon contributions, the magnetic specific heat exhibits a broad Schottky anomaly at $\sim$40~K. The peak can be nicely described by a two-level Schottky model as
$$C_m(T) = R(\frac{\Delta}{T})^2\frac{\mathrm{exp}(\Delta/T)}{[1+\mathrm{exp}(\Delta/T)]^2},$$
where $R$ is the ideal gas constant. The best fit yields a gap ($\Delta$) of 8.55(2) meV, consistent with the INS measurements. The successful application of the two-level function also suggests that the level at $\sim$8.55 meV is doubly degenerate.

\section{ac susceptibility}

\begin{figure*}
\includegraphics[width=0.95\textwidth]{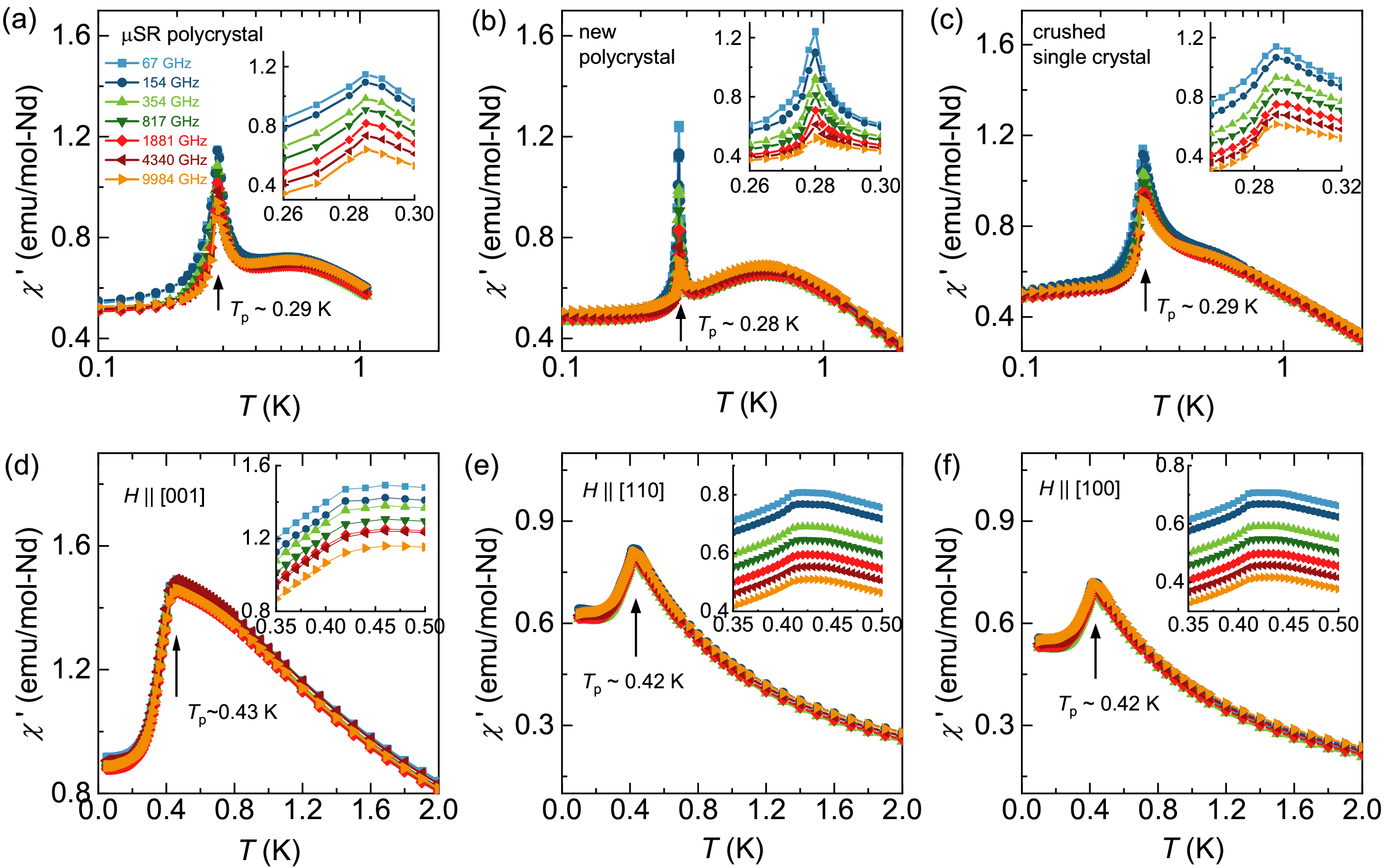}
\caption{AC susceptibility measurements on (a) the polycrystal used for the $\mu$SR experiment, (b) a newly synthesized polycrystal, (c) a crushed single crystal, and (d-f) a single crystal along different crystallographic directions. The insets display enlarged views of the region around the peak. The data have been shifted vertically for clarity.}
\label{ac}
\end{figure*}
To rule out the possibility of a spin-glass state at low temperatures, we performed ac susceptibility measurements on the polycrystalline samples used for the $\mu$SR experiments. In contrast to the single crystal data as shown in our previous paper \cite{Liu20242}, a sharp peak is observed at $\sim$0.29 K, followed by a broad peak at around 0.6 K; see Fig. \ref{ac}(a). This behavior, however, is consistent with the specific heat results and reproducible; compare to Fig. \ref{Cp} and Fig. \ref{ac}(b), respectively. To elucidate the origin of this low temperature anomaly, we performed additional single crystal measurements with magnetic field applied along the crystallographic [001], [110], and [100] directions, as shown in Fig. \ref{ac}(d-f). All these measurements show a consistent kink at around 0.42 K. We thus crushed the single crystal, which was used for Fig. \ref{ac}(e), into powders and performed the measurement again. As can be seen in Fig. \ref{ac}(c), the sharp peak at 0.29 K appears, and the overall behavior is consistent with that in Fig. \ref{ac}(a). We thus speculate that the peak originates from intrinsic magnetic ordering in which the moments prefer to lie within the mirror plane as revealed by our neutron diffraction measurements, rather than from extrinsic effects such as defects in the polycrystalline samples. The exact direction is difficult to be aligned due to the geometry of the sample and the sample holder of our dilution refrigerator. Having established the transitions, we focus on the insets of Fig. \ref{ac} which display the enlarged views of the region around the peak. Clearly, no discernible shift of the peak positions, either at 0.29~K, or at 0.42~K, could be observed, ruling out the formation of a glassy state.

\section{ESR}
\begin{figure*}
\includegraphics[width=0.95\textwidth]{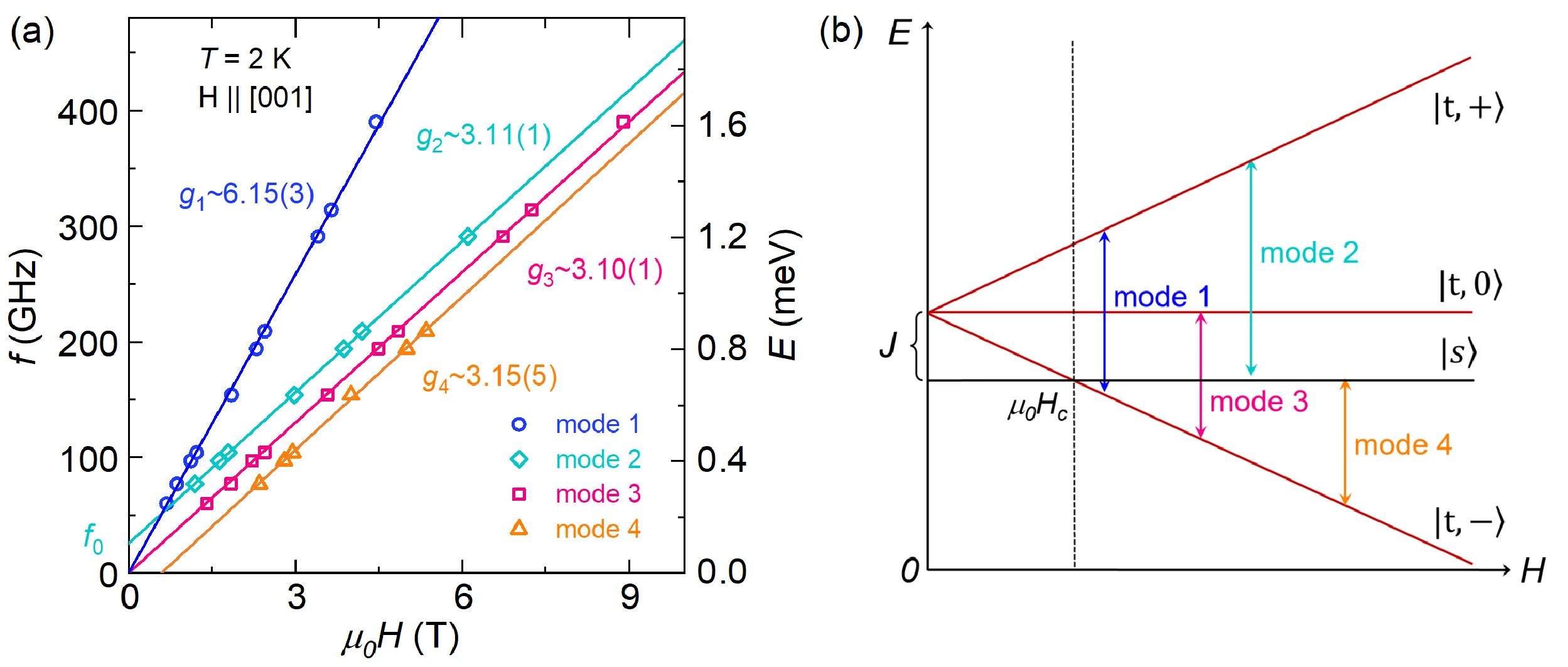}
\caption{Electron spin resonance measurements. (a) Resonance frequencies as a function of magnetic field applied along the $c$ axis, together with linear fits. (b) A schematic of the dimer model showing the corresponding excitations observed in (a).}
\label{esr}
\end{figure*}
Figure \ref{esr}(a) shows the electron spin resonance (ESR) spectra measured at 2~K with magnetic field applied along the \textit{c} axis. Four modes can be identified. Linear fits to the data yield three \textit{g} factors ($g_2 - g_4$) of about 3.1, whereas $g_1$ is approximately doubled. Moreover, linear extrapolations show that modes 1 and 3 pass through the origin, while mode 2 exhibits a zero-field gap and mode 4 has a level crossing at about 0.59~T. Such behavior can be naturally explained by an antiferromagnetic Heisenberg dimer model as shown in Fig. \ref{esr}(b). In such a case, the zero-field ground state is a singlet ($|00\rangle$), which is separated from the triplet ($|10\rangle, |11\rangle, |1-1\rangle$) by the energy \textit{J}. Singlet-triplet transitions, as well as the $\Delta S_z$ = $\pm$2 transitions within the triplet are usually forbidden in ESR measurements. The observed transitions thus indicate a mixing among the eigenstates for the pure Heisenberg model such as $|s\rangle = \alpha|00\rangle + \beta|10\rangle$, which enables mode 2 and mode 4. Symmetry analysis reveals that a DM vector perpendicular to the dimer bond is allowed \cite{Toth_2015}, which accounts for the observed mixing. From the observed zero field gap, the intradimer interaction energy $J$ is estimated to be 0.106(2) meV, which is also consistent with that estimated from $g_4\mu_B\mu_0 H_c$ = 0.11(1) meV.

\section{CEF analysis}

\begin{figure*}
\includegraphics[width=0.95\textwidth]{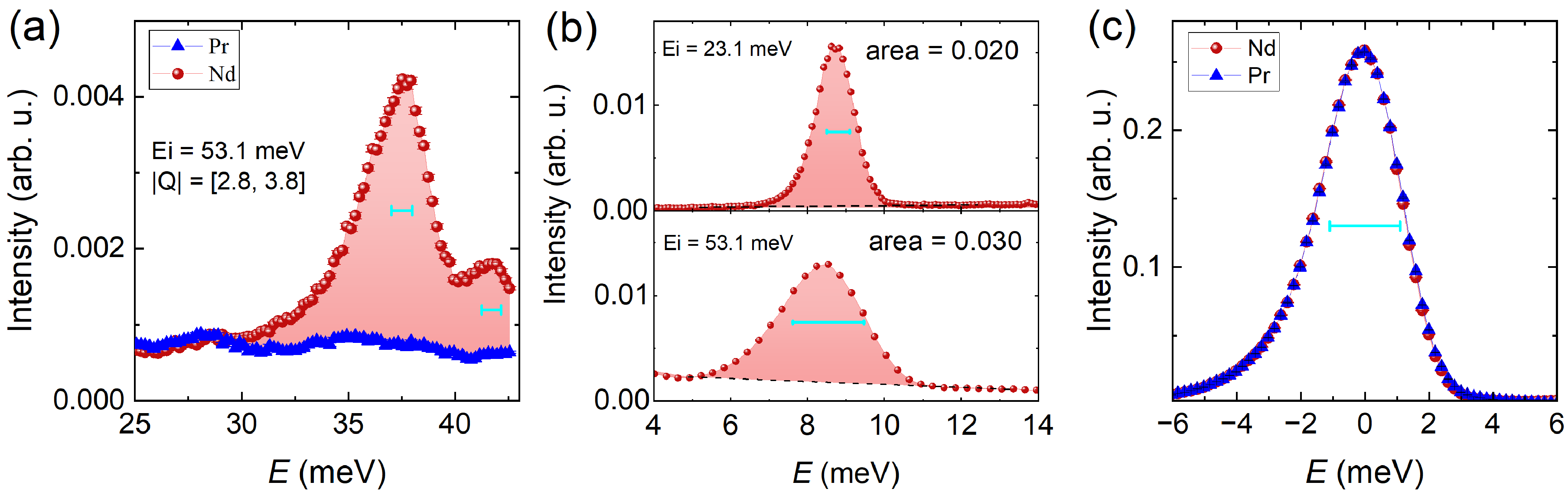}
\caption{Inelastic neutron scattering spectra. (a) Comparison of the energy cuts for the Pr and Nd compounds. (b) Comparison of the integrated intensities with different Ei's for the Nd compound. (c) Elastic lines for the Pr and Nd compounds. The horizontal bars represent the instrument resolutions computed from the PyChop software.}
\label{INS2}
\end{figure*}

Figure \ref{INS2}(a) compares the energy cuts of the INS spectra for the \NBGO\ and \PBGO\ compounds. Apparently, there is no significant phonon contributions above 30 meV as seen from the Pr data. Therefore, the observed intensities for the \NBGO\ compound should have a magnetic origin. The observed peaks are substantially broader than the instrument resolution, especially on the low energy side of the peak centered at $\sim$37.5 meV. Note that the calculated peak widths from PyChop underestimate the actual instrument resolution, as indicated by the elastic line width shown in Fig.~\ref{INS2}(c). Even after accounting for this effect, the observed peak is still too broad to be attributed to a single peak, suggesting the overlap of multiple peaks in this energy range.

Since the 8.7-meV peak shape is strongly influenced by the elastic line with $E_i$ = 53.1 meV, we took the peak with $E_i$ = 25.1 meV for CEF analysis. To normalize the intensities obtained from different $E_i$s, a linear background was subtracted from both datasets, and the area below the peak was integrated without fittings, as shown in Fig. \ref{INS2}(b). A ratio of 1.5 was then used for subsequent CEF fittings.

Based on the obtained CEF model, the calculated \textit{g} tensor is
\[ g = \begin{pmatrix}
  1.17 & 0 & -1.64 \\
  0 & 0.14 & 0 \\
  -0.68 & 0 & 3.10 \\
\end{pmatrix}.\]
The \textit{g} factor along an arbitrary  direction \textbf{n} can then be calculated as
$g_n = \sqrt{\textbf{n}^{T}gg^{T}\textbf{n}}$, where the superscript T represents the transpose of the matrix \cite{ESR}. Accordingly, the \textit{g} factors along the local \textit{x}, \textit{y} and \textit{z} axes are 2.02, 0.14 and 3.18, respectively.

\section{Low temperature magnetization}
\begin{figure}
\includegraphics[width=0.45\textwidth]{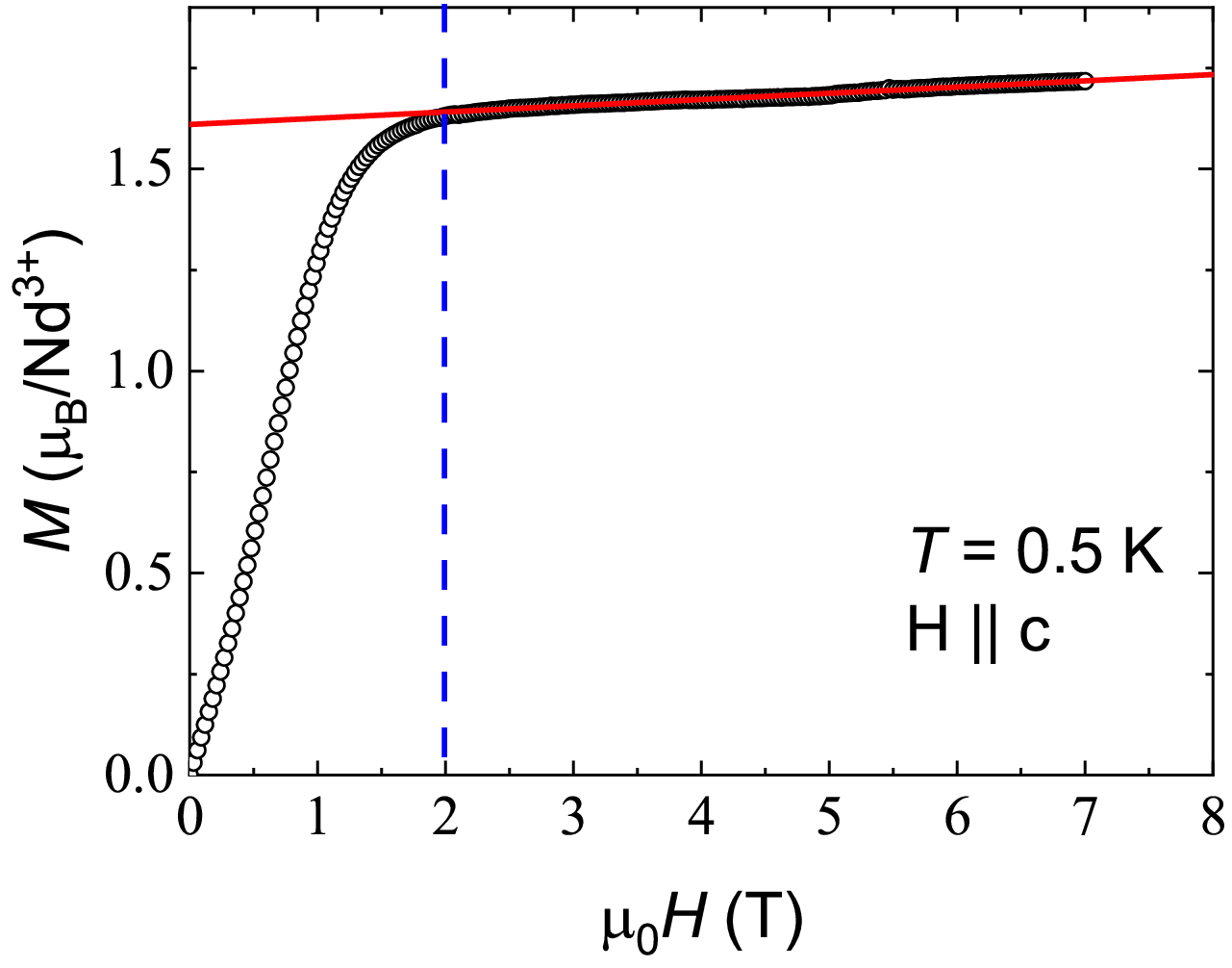}
\caption{M vs. H at 0.5 K.}
\label{MH}
\end{figure}

Figure \ref{MH} shows the isothermal magnetization measured at 0.5 K. A linear fit above 3~T yields a saturation moment of 1.61 $\mu_\mathrm{B}$/Nd$^{3+}$. This is consistent with the CEF fit which estimates a saturation moment of $g_z$/2 = 1.59 $\mu_\mathrm{B}$/Nd$^{3+}$. The fit also demonstrates that the saturation field is $\sim$2~T at 0.5 K.

\begin{table*}
\caption{CEF eigenvalues and eigenvectors ($|Jm_J\rangle$ basis) for \NBGO. The CEF parameters are $B_2^0$~=~0.2148(2), $B_2^1$~=~-0.5803(1), $B_2^2$~=~-0.5694(3), $B_4^0$~=~0.0020(1), $B_4^1$~=~-0.0288(1), $B_4^2$~=~-0.0048(3), $B_4^3$~=~0.01587(2), $B_4^4$~=~-0.0015(2), $B_6^0$~=~-0.000115(2), $B_6^1$~=~0.00181(3), $B_6^2$~=~-0.00119(3), $B_6^3$~=~-0.00054(3), $B_6^4$~=~0.00062(1), $B_6^5$~=~0.001(1), $B_6^6$~=~0.00005(4)~meV.}
\begin{tabular}{c|cccccccccc}
E (meV) &$| -\frac{9}{2}\rangle$ & $| -\frac{7}{2}\rangle$ & $| -\frac{5}{2}\rangle$ & $| -\frac{3}{2}\rangle$ & $| -\frac{1}{2}\rangle$ & $| \frac{1}{2}\rangle$ & $| \frac{3}{2}\rangle$ & $| \frac{5}{2}\rangle$ & $| \frac{7}{2}\rangle$ & $| \frac{9}{2}\rangle$ \tabularnewline
 \hline
0.000 & 0.0 & 0.0009 & -0.126 & 0.0142 & -0.2744 & 0.1868 & -0.4818 & 0.6863 & 0.3135 & 0.2688 \tabularnewline
0.000 & -0.2688 & 0.3135 & -0.6863 & -0.4818 & -0.1868 & -0.2744 & -0.0142 & -0.126 & -0.0009 & 0.0 \tabularnewline
8.706 & -0.3234 & 0.3305 & -0.3843 & 0.6086 & 0.4379 & 0.2339 & 0.0976 & 0.1082 & -0.0107 & -0.0176 \tabularnewline
8.706 & 0.0176 & -0.0107 & -0.1082 & 0.0976 & -0.2339 & 0.4379 & -0.6086 & -0.3843 & -0.3305 & -0.3234 \tabularnewline
35.980 & 0.2569 & -0.0303 & -0.2093 & 0.1576 & -0.4053 & 0.0785 & 0.3882 & 0.3044 & -0.6613 & 0.1155 \tabularnewline
35.980 & -0.1155 & -0.6613 & -0.3044 & 0.3882 & -0.0785 & -0.4053 & -0.1576 & -0.2093 & 0.0303 & 0.2569 \tabularnewline
37.451 & 0.472 & 0.2353 & -0.1651 & 0.3018 & -0.4006 & 0.0948 & 0.2166 & -0.3171 & 0.5354 & 0.0045 \tabularnewline
37.451 & 0.0045 & -0.5354 & -0.3171 & -0.2166 & 0.0948 & 0.4006 & 0.3018 & 0.1651 & 0.2353 & -0.472 \tabularnewline
41.463 & 0.7216 & 0.0191 & -0.2898 & -0.1003 & 0.5263 & -0.1629 & -0.2531 & 0.0741 & -0.1085 & 0.0034 \tabularnewline
41.463 & 0.0034 & 0.1085 & 0.0741 & 0.2531 & -0.1629 & -0.5263 & -0.1003 & 0.2898 & 0.0191 & -0.7216 \tabularnewline
\end{tabular}
\label{EV}
\end{table*}

\begin{table}
\caption{Irreducible representations (IRs) together with basis vectors $\psi_i$ for Nd ions at the 4\textit{e} site [Nd1(0.16, 0.34, 0.503), Nd2(0.84, 0.66, 0.503), Nd3(0.34, 0.84, 0.497), Nd4(0.66, 0.16, 0.497)] with the space group $P\bar{4}2_1m$ and propagation vector \textbf{k}~=~(0, 0, 0). \label{BV}}
\begin{tabular}{llllll}
IRs & $\psi_i$  & Nd1  & Nd2  & Nd3  & Nd4 \\
 \hline
$\Gamma_1$ & $\psi_1$  & (1~1~0) & (-1~-1~0) & (-1~1~0)& (1~-1~0)\\
$\Gamma_2$ & $\psi_2$  & (1~-1~0) & (-1~1~0) & (-1~-1~0) & (1~1~0) \\
           & $\psi_3$  & (0~0~1) & (0~0~1) & (0~0~-1)& (0~0~-1) \\
$\Gamma_3$ & $\psi_4$  & (1~-1~0) & (-1~1~0) & (1~1~0) & (-1~-1~0) \\
           & $\psi_5$  & (0~0~1) & (0~0~1) & (0~0~1)& (0~0~1) \\
$\Gamma_4$ & $\psi_6$  & (1~1~0) & (-1~-1~0) & (1~-1~0)& (-1~1~0)\\
$\Gamma_5$ & $\psi_7$  & (1~0~0) & (1~0~0)  & (-1~0~0)  & (-1~0~0)  \\
           & $\psi_8$  & (0~1~0)  & (0~1~0) & (0~1~0)& (0~1~0) \\
           & $\psi_9$  & (0~0~1) & (0~0~-1) & (0~0~-1)& (0~0~1) \\
           & $\psi_{10}$& (0~-1~0)  & (0~-1~0) & (0~1~0)& (0~1~0)  \\
           & $\psi_{11}$  & (-1~0~0) & (-1~0~0)  & (-1~0~0)  & (-1~0~0) \\
           & $\psi_{12}$ & (0~0~1) & (0~0~-1) & (0~0~1)& (0~0~-1) \\
\end{tabular}
\end{table}

\end{document}